\documentclass[universe,review,accept,oneauthor,pdftex]{Definitions/mdpi} 

\usepackage{natbib}

\newcommand\aap{A\&A}                
\newcommand\aapr{A\&ARv}             
\newcommand\aj{AJ}                   
\newcommand\apj{ApJ}                 
\newcommand\apjl{ApJ}                
\newcommand\araa{ARA\&A}             
\newcommand\jcap{J.~Cosmol. Astropart. Phys.} 
\newcommand\mnras{MNRAS}             
\newcommand\nar{New~Astron.~Rev.}    
\newcommand\nat{Nature}              
\newcommand\prd{Phys. Rev.~D}        
\newcommand\prl{Phys. Rev.~Lett.}    
\newcommand\pasj{PASJ}               
\newcommand\ssr{Space Sci. Rev.}     

\firstpage{1} 
\pubvolume{1}
\issuenum{1}
\articlenumber{0}
\pubyear{2021}
\copyrightyear{2021}
\externaleditor{Academic Editor: Ulisses Barres de Almeida, Michele Doro} 
\datereceived{26 July 2021} 
\dateaccepted{26 August 2021} 
\datepublished{} 
\hreflink{https://doi.org/} 

\Title{The Hunt for Pevatrons: The Case of Supernova Remnants}

\TitleCitation{The Hunt for Pevatrons: The Case of Supernova Remnants}

\Author{{Pierre Cristofari} 
 \orcidA{}}

\AuthorNames{Pierre Cristofari}

\AuthorCitation{Cristofari, P.}

\address[1]{%
Observatoire de Paris, LUTH, 5 Place Jules Jansen, 92195 Meudon, France; pierre.cristofari@obspm.fr\\
}

\abstract{The search for Galactic pevatrons is now a well-identified key science project of all instruments operating in the very-high-energy domain. Indeed, in this energy range, the detection of gamma rays clearly indicates that efficient particle acceleration is taking place, and observations can thus help identify which astrophysical sources can energize particles up to the $\sim$PeV range, thus being $pevatrons$. In the search for the origin of Galactic cosmic rays (CRs), the PeV range is an important milestone, since the sources of Galactic CRs are expected to accelerate PeV particles. This is how the central scientific goal that is 'solving the mystery of the origin of CRs' has often been distorted into 'finding (a) pevatron(s)'. Since supernova remnants (SNRs) are often cited as the most likely candidates for the origin of CRs, 'finding (a) pevatron(s)' has often become 'confirming that SNRs are pevatrons'. Pleasingly, the first detection(s) of pevatron(s) were not associated to SNRs. Moreover, all clearly detected SNRs have yet revealed to not be pevatrons, and the detection from VHE gamma rays from regions unassociated with SNRs, are reminding us that other astrophysical sites might well be pevatrons.
This short review aims at highlighting a few important results on the search for Galactic pevatrons.}


\begin{document}
\section{Introduction}
\subsection{The Hunt for the Sources of Galactic Cosmic Rays}

The term \textit{Pevatron}, simply referring to an object capable of accelerating particles up to the PeV (=$10^{15}$ eV) range, is now widely used in the scientific programs of all major instruments investigating particle acceleration in astrophysical sources. This passionate search for pevatrons is not a stubborn search for the most powerful astrophysical particle accelerators, but is undoubtedly motivated by the hunt for the origin of Galactic cosmic rays (CRs).

CRs are relativistic charged particles, typically $\sim$92\% protons, $\sim$6\% Helium, $\sim$1\% electrons and $\sim$1\% heavier nuclei, that fill the whole disk of the Milky Way. Although decades of experiments and theoretical studies helped to accumulate valuable knowledge, the fundamental question of their origin is still missing~\citep{gaisser1990}. 

It is widely accepted that accelerators located inside our Galaxy must produce the bulk of CRs detected at the Earth. One strong argument for the Galactic origin comes from the detection of gamma rays: observations of the Galaxy revealed that the Galactic disk is gamma-ray bright, which is a clear indication of the interactions of CRs with interstallar medium (ISM) material and Galactic magnetic field in the disk, and that gamma rays roughly scale with the amount of material in the disk. Moreover, observations of the Large Magellanic Cloud (LMC), located in the vicinity of our Galaxy at $\sim$50 kpc, have shown that the gamma-ray signal is weaker than expected given the mass of the target material in the LMC~\citep{ackermann2016,tibaldo2021}. This reduced gamma-ray signal from the LMC can be seen as evidence that CRs are not homogeneously distributed in the Universe, and originate within our Galaxy.
Additionally, a simple reasoning on the Larmor radius $r_L$ of CRs, compared to the typical Galactic halo $H$ size imposes that CRs of energy $E$ remain confined within the Galaxy if $r_L(E) \lesssim H$. For typical magnetic fields of the interstellar medium (ISM) of the order of $\sim$$\upmu$G and a halo size of a few $\sim$kpc (although both of these quantities remain poorly constrained), this simple estimate indicates that CR protons of energy $E \lesssim 10^{16-17}$eV are expected to remain confined in the Galaxy. Therefore, we have indications that (1) CRs originate inside our Galaxy; (2) CRs are confined inside our Galaxy up to at least $\lesssim 10^{16}$--$10^{17}$eV. {These CRs of Galactic origin are simply called Galactic CRs.}

Moreover, the local CR spectrum is famously exhibiting a power law of spectrum $\propto E^{-2.7}$ up to an energy of $\sim$3 PeV, where the slope becomes $\propto E^{-3}$, this feature being usually referred to as the \textit{knee}~\citep{antoni2005}. Let us here mention that some collaborations have also found that the \textit{knee} of the light component ($He$+protons) might be located at lower energy around $\sim$0.7 PeV~\citep{bartoli2015,cao2020}, which could mean that the proton knee is below 1 PeV. 

The remarkable, and almost featureless power law (although to be precise several instruments have in fact revealed some deviations in the powerlaw~\citep{dampe_protons}), up to the knee, is seen as an indication that the sources of Galactic CRs must be accelerating particle up to the knee, and therefore, be $\textit{pevatrons}$. Hence, the wild hunt for Galactic pevatrons, with a sometimes oversimplified argument that ``finding the pevatrons'' would mean ``unveiling the origin of CRs up to the knee'', or ``unveileing the origin of Galactic CRs''. Let us here insist on the fact that for the source(s) of Galactic CRs, being (a) pevatron(s) is a necessary condition but is not sufficient. It is for instance clear now that the Crab nebula is an astrophysical site where the acceleration of PeV particle (at least electrons) is taking place, and is however, not currently accepted as a typical source of the bulk of Galactic CRs (protons).

\subsection{100 TeV Gamma Rays}

The best way to probe particle acceleration up to the PeV range, is (so far) through gamma-ray observations, in the very-high-energy range (VHE, 100 GeV to 100 TeV), and ultrahigh-energy range (UHE, above 100 TeV). Indeed, radiations in the VHE and UHE range directly testify to the abundant presence of nonthermal particles. Three mechanisms can essentially produce gamma rays in this range. First, a hadronic mechanism: the creation and decay of neutral pions through the interactions of accelerated hadrons with nuclei of the ISM. Second, a leptonic mechanism: the inverse Compton scattering (ICS) of accelerated electrons on soft photons (CMB, optical, infrared). Third, another leptonic mechanism: the Bremsstrahlung emission of accelerated electrons that can also lead to the production of gamma rays in this range~\citep{longair,hinton2009}.

In the $\sim$TeV domain, it is not easy to differentiate between these different potential origins, and to know if parent hadrons or leptons are responsible for the production of gamma rays. The interpretation of the origin of observed TeV gamma rays is often subject to fervent discussions, and in many cases, the gamma-ray emission can be explained by one mechanism or another. See e.g., the numerous discussions on many well-studied astrophysical objects, for which the GeV and TeV gamma-ray emission can often be accounted for by a leptonic ICS origin, a hadronic origin, or a mixture of both. The case of the superova remnant (SNR) RXJ1713-3946 is a stereotypical example of the copious discussions on the interpretation of the gamma-ray emission.~\citep{berezhko2008,morlino2009,fang2009,casanova2010,zirakashvili2010,ellison2012,dermer2013,yang2013,kuznetsova2019,tsuji2019,zhang2019,fukui2021,cristofari2021b}

It has often been claimed that above $\gtrsim$50 TeV, the interpretation of detected gamma rays becomes less ambiguous, since the inverse Compton scattering of nonthermal electrons becomes dramatically inefficient in producing gamma rays, due to the Klein--Nishina suppresion: at these energies, gamma rays are thus expected to testify that the acceleration of hadrons is taking place. Moreover, the VHE gamma rays of hadronic origin mimic the parent proton distribution, with a typical energy of gamma rays scaling as $E_{\gamma} \approx 1/10 E_{p}$. Therefore, this has often been condensed into ``the detection of $\sim$100 TeV gamma rays is be a direct indication of acceleration of $\sim$PeV CR protons''.

Still, this affirmation needs to be tempered. The distinction between hadronic and leptonic origin of gamma rays is far from being straightforward even in the $\sim$100 TeV range. Although the Klein--Nishina suppression considerably reduces the production of gamma rays from ICS in this range, leading to an exponential suppression above \mbox{$\gtrsim$50 TeV}, gamma rays can still be produced by electrons through ICS. A conspicuous amount of VHE electrons can thus produce a gamma-ray signal in the 100 TeV region. Such a gamma-ray signal can especially be important in the context of instruments operating with an ever increasing sensitivity in the $\gtrsim$50 TeV range, hence capable of detecting gamma rays photon even from steep or exponentially suppressed spectra~\citep{bai2019,CTAscience,SWGO}.
Moreover, the synchrotron emission of accelerated electrons can also efficiently produce gamma rays. 

A prime example of a gamma-ray emission interpreted as the result of accelerated electrons is the Crab Nebula, the brightest source of the TeV sky, and thus one of the best studied objects~\citep{balbo2011,aharonian2020,cao2021}. More generally, several theoretical works have illustrated that electrons accelerated above $\gtrsim$100 TeV could lead, through inverse Compton scattering or bremsstrahlung, to gamma rays in the $\gtrsim$100 TeV range~(see e.g., \citep{hinton2009} and references therein), and can produce hard gamma-ray spectra extending up to the $\gtrsim 100$ TeV, making it especially strenuous to differentiate between hadronic and leptonic acceleration mechanisms~\citep{brehaus2021}. Let us additionnally mention that the detection of neutrinos would be an effective and elegant way to discriminate between the hadronic and leptonic mechanisms since neutrinos can only be produced through hadronic interactions~\citep{anchordoqui2014}.

The detection and study of electron accelerators is of course of great importance, but in the search for the origin of CRs, is not enough. First, because CRs are mostly protons, and we are thus looking for proton sources: an efficient electron accelerator is not necesseraliy an efficient proton accelerator. The case of SNRs is instructive, since in spite of extensive studies the content of electrons and protons accelerated at SNRs is still poorly understood. Second, because unlike protons, electrons suffer important losses while transported in the Galaxy. Therefore, the problem of the origin of CR electrons is more \textit{local} than the one of CR protons and somewhat disconnected~\citep{amato2018,aguilar2019,evoli2020,evoli2021}.

So far, on the observational side, all major observatories relying on Imaging Atmospheric Cherenkov Telescopes (IACTs), such as VERITAS~\citep{VERITAS}, MAGIC~\citep{albert2008}, or H.E.S.S.~\citep{aharonian2006}, and predecessors, such as HEGRA~\citep{aharonian2004}, have been probing VHE domain up to a few tens of TeV. Therefore, the search and identification of pevatrons has been a remarkably challenging task, but nonetheless successful since H.E.S.S. was capable of claiming the detection of the first pevatron in the Galactic center region~\citep{HESSpevatron}. Other observatories, using different techniques, relying on shower front detectors (or ``air shower detectors''), such as HAWC~\citep{abeysekara2019}, Tibet AS$\gamma$~\citep{Tibet2019} and LHAASO~\citep{cao2021}, operating up to and above the $\sim$100 TeV gamma-ray domain also reported on the detection of several Galactic pevatrons. All these observations indicated that most of the pevatron candidates seem to not be associated to SNRs.

In this short review, we intend to especially discuss the case of SNRs, since they have been, for long, at the center of the debate on the origin of Galactic CRs, and that the question of whether they can accelerate PeV particles is still open. We then briefly discuss other pevatron candidates and important open questions in the coming years. For pedagogical introduction to the general physics of CRs, we refer the reader to reference writings of the field~\citep{ginzburg1964,gaisser1990,berezinskii1990}. For updated reviews discussing recent results and advances on the problem of the origin of CRs, we refer the reader to~\citet{blasi2019,gabici2019}.



\section{Supernova Remnants}
In the search for the origin of Galactic CRs, supernova remnants (SNRs), the spherical shock waves expanding in the ISM after the explosion of massive stars, have become over the years the most famous candidates. Several reviews detail the reasons of the success of the SNR paradigm~\citep{drury1983,blasi2013,amato2014}. In short, the SNR hypothesis is supported by at least the following strong arguments: 
\begin{enumerate}
\item Energetic considerations: a simple back-of-the-envelope calculation illustrates that converting a fraction ($\sim$1\%--10\%) of the total explosion energy of SNe into CRs can account the measured energy density of CRs at the Earth. 

\item Spectrum of accelerated particles: at SNRs, which are strong shock waves (i.e., of compression factor $\gtrsim$4), a first order \textit{Fermi} mechanism, referred to as \textit{Diffusive shock acceleration} (DSA)~\citep{axford1977,krymskii1977,bell1978,bell1978_2} can explain the efficient acceleration of particles with a power law in momentum $\propto p^{-\alpha}$, with $\alpha \sim 4$ for a strong shock (which corresponds, at high energy to $E^{-\alpha  +2}$), somewhat compatible with the Local CR spectrum measured at the Earth. 

\item Magnetic field amplification: several mechanisms have been shown to be able to amplify the magnetic field at SNRs shock fronts, therefore helping to the reach the PeV range. 
Indeed, the Hillas criterion, where the Larmor radius of particles $r_L$ is equated to the typical size of the accelerator (the SNR radius $r_{\rm sh}$) gives that the maximum energy of particles is typically at most: 
\begin{equation}
E_{\rm max}\approx  \left( \frac{r_{\rm sh}}{\text{pc}} \right) \left( \frac{u_{\rm sh}}{\text{1000 km/s}} \right) \left( \frac{B}{\mu\text{G}}\right) \text{TeV}
\end{equation}
with $u_{\rm sh}$ the SNR shock velocity, and $B$ the magnetic field. Therefore, in order to attain the PeV range, for a typical SNRs of a few pc, expanding at a few 1000 km/s, values of at least $\gtrsim 10^{2}$ $\upmu$G are needed. More precisely, there are now established theoretical results on the growth of instabilities at colisionless shocks, coupled to observational X-ray measurements at SNR shocks, that have reported values of magnetic fields of several hundreds of $\upmu$G, i.e., several orders of magnitude above the values of the ISM~\citep{vink2012}.

\item Gamma rays: the detection of GeV to multi-TeV gamma rays from SNRs, and especially the capacity of current IACTs to resolve the shells corresponding to the spherical expanding shock waves, is an undeniable indication that efficient particle acceleration is taking place. 
\end{enumerate}

In spite of these compelling arguments, the SNR hypothesis is however facing several difficulties, discussed in detail in several reviews~\citep{tatischeff2018,gabici2019}. Amongst them: 
\begin{enumerate}
\item Gamma-ray spectra: in the TeV range, the spectral index of all SNRs appears to be $>$2. This is problematic since the gamma-ray spectrum is expected to mimic the spectrum of particles accelerated at the SNR shock. Hence, this seems to indicate that the spectrum of particles accelerated at the shock is steeper than what is expected in the test particle limit $\propto E^{-2}$. Although the authors of various works have studied how efficient CRs at shocks can modified the produced spectrum, though nonlinear effects~\citep{malkov1997,amato2005,amato2006,caprioli2012,caprioli2020,haggerty2020}, the question of the slope of particles accelerated at SNR shocks is still open. 

 \item The spectrum of particles injected in the ISM: from B/C measurements, the effects of the transport in the Galaxy have been estimated, finding that it should affect the CR spectrum by $\propto E^{-0.3\dots0.6}$~(see e.g., \citep{amato2021}). In order to produce the spectrum of the local CR spectrum below the knee $\propto E^{-2.7}$, this means that the CR sources must inject in the ISM a spectrum $\propto E^{-2.1\dots2.4}$, i.e., steeper than what is predicted in the usual test particle limit ($\propto E^{-2}$). The problem of the spectrum injected by SNRs in the ISM is still open~\citep{diesing2019,cristofari2021}. 

\item The $^{22}$Ne/$^{20}$Ne abundance measured in CRs is a factor $\sim$5 larger than the one in the solar wind~\citep{higdon2003,binns2006}. This overabundance is not easily accounted for under the SNR hypothesis alone~\citep{prantzos2012}. This has motivated the study of alternative scenarios, in which particles are at least in part accelerated in an environment enriched by stellar outflows (e.g., massive stars, stellar clusters, OB associations, superbubbles) and might explain the overabundance of $^{22}$Ne/$^{20}$Ne~\citep{gupta2020}. Other issues in the chemical composition of CRs have also been reported: for example, the enhanced $^{58}$Fe/$^{56}$Fe ratio, or the linear dependency of the B and Be abundance with metallicity in metal poor halo stars, that are not easily explained under the SNR hypothesis~\citep{tatischeff2018}.

\item No pevatron: As mentioned in the introduction, the sources of CRs are expected to be pevatrons. This means that SNRs are expected, for at least a period of their evolution, to be pevatrons. Since the gamma rays from accelerated protons scale as $ E_{\gamma } \propto 1/10 E_{p} $, the detection of gamma rays in the $\sim$100 TeV range probes the acceleration of PeV CR protons. So far, the observation of all SNR shells in the VHE domain have revealed cut offs indicating that PeV CRs are not efficiently produced. 
\end{enumerate}

This last point is crucial, since of all the aspects mentioned above, the identification of a SNR pevatron is often seen to be the conclusive proof that SNRs are the sources of Galactic CRs. Hence, this has stimulated the investigation of alternative scenarios for the origin of CRs. At least two comments are in order here. First, stricto census, current IACTs are not sensitive enough to detect gamma rays in $\sim$100 TeV range, but they still have been able to measure the gamma-ray spectrum in the TeV and tens of TeV region, and revealing exponential suppression that indicate that the flux in the 100 TeV range should be at least drastically suppressed. Second, recent results of several instruments, HAWC, TibetAS$\gamma$ and LHAASO~\citep{albert2020,2021tibet,cao2021} have reported on the detection of $\sim$100 TeV gamma rays from a region in which the SNR G106.3+2.7 is located. By design, the improved sensitivity in the 100 TeV range of these observatories, compared to typical IACTs, comes with a poorer angular resolution than typical IACTs. It is therefore difficult for these instruments to spatially resolve the source(s) of the 100 TeV emission and specify which astrophysical object(s) is (are) responsible for the acceleration of particles up to the PeV range~\citep{albert2020}. In addition, the associated SNR, G106.3+2.7 is middle-aged ($\sim$10 kyr) which makes it a priori a relatively poor pevatron candidate~\citep{acciari2009}. The role of the different objects located in this complex region have thus yet to be clarified.

Two possible explanations for the fact that all known TeV shells seem to not be pevatrons are: either that SNRs are only pevatrons for a short period of time, a priori in the early stages of their evolution and that all studied SNRs are already too old to accelerate PeV particles, or/and, that only a small fraction of SNRs are pevatrons~\citep{cristofari2018,cristofari2020}. In both cases, this coud explain the limited chances to identify an active SNR pevatron, even in the future Galactic surveys.

\subsection{Acceleration up to the PeV Range}

One of the reasons why all best studied SNRs are not pevatrons might be because they are too old to still be pevatrons. Indeed, as a SNR blast wave expands in the ISM, it slows down. The maximum energy of accelerated particles is thus expected to decrease. Naively, if we estimate that the maximum energy can be estimated by equating the diffusion length of particles $l_{\rm d}$ to a fraction $\chi$ of the shock radius $r_{\rm sh}$: 
\begin{equation}
l_{\rm d}= \frac{D (E)}{u_{\rm sh}(t)} \approx \chi r_{\rm sh} (t)
\end{equation}
assuming a Bohm-like diffusion coefficient, for relativistic CR particles, $D(E) = \frac{1}{3} r_{\rm L}(E) c$, with $B\propto t^{-\beta}$, $r_{\rm sh} \propto t^{\alpha}$, and $u_{\rm sh} \propto t^{\alpha -1}$, this leads to: 
\begin{equation}
E_{\rm max} (t) \propto B(t) r_{\rm sh}(t) u_{\rm sh}(t) \propto t^{2\alpha -1 - \beta}
\end{equation}
thus $E_{\rm max}$ decreases in time provided that $\beta > 2 \alpha -1$. For a typical SNR expanding in a uniform ISM, $\alpha = 4/7$~\citep{chevalier1982,chevalier1994} thus $E_{\rm max}$ decreases if $\beta \geq 1/7$. For a SNR expanding in a wind $\propto r^{-2}$, $\alpha=7/8$, $\beta \geq 3/4$. 

The nature of the mechanism driving the amplification of the magnetic field, needed to reach the PeV range is currently still a matter of debate. It has been shown that several mechanisms can theoretically lead to the growth of instabilities. Let us for example mention~see also the short review \citep{gabici2016}: 
\begin{enumerate}
\item Acoustic instabilities, where a gradient of the CR pressure $\nabla P$ in a fluid of density $\rho$ leads to the acceleration of the fluid $\propto \nabla P/ \rho$. In an inhomogeneous medium, the inhomogeneities of density will in turn lead to inhomogeneities in the acceleration, therefore amplifying the starting density fluctuations~\citep{drury1986,drury2012,downes2014,delvalle2016}. 

\item Density fluctuations in turbulent magnetized plasma, where instabilities are amplified by the dynamics of the shock (and where CRs do not play any role in the growth of instabilities): the production of PeV CRs by this mechanism is however expected to be rather limited~\citep{giacalone2007,guo2012}.

\item Resonant streaming instabilities: induced by accelerated CRs, the background plasma reacts to moving CRs and creates a current compensating the positive charge excess created by CRs. For particle of Larmor radius $r_L$, this leads to the growth of \textit{resonant} waves of wave number $k \approx 1/r_{L}$. The resonant nature of this process typically leads to a saturation when the amplified field $\delta B$ becomes of the order of the pre-existing ordered magnetic field $B_0$.~\citep{skilling1975,achterberg1983}

\item Nonresonant streaming instabilities (often referred to as \text{Bell} instability): as in the resonant streaming instability, the current of CRs $\vec{j}_{\rm CR} $ accelerated at the shock cause the background to react with an oppositely directed current. This results in a force $\vec{j}_{\rm CR} \times \vec{B}$ responsible for the growth of perturbations in the magnetic field. The growth is the fastest on scales much smaller than the Larmor radius of particles, hence the nonresonant label~\citep{bell2004,bell2013,schure2013,schure2014}.

\end{enumerate}

This so-called nonresonant Bell mechanism is especially important in the case of SNRs, because it can lead to magnetic field amplification with values significantly larger than the pre-existing magnetic field $B_0$. Moreover, the growth rate of the instabilities is typically faster than in all previously mentioned mechanisms, and is therefore thought to be the mechanism governing magnetic field amplification at SNRs. A pedagogical discussion on the growth of resonant and nonresonant streaming instabilities can be found in~\citep{blasi2019}. 
The growth of the magnetic field from nonresonant streaming instability is exponential until saturation is reached. The exact details on when saturation is reached is still a matter of discussion, but the typical amplified magnetic field $\delta B$ can be estimated equating the Larmor radius of particles to the typical plasma spatial displacement induced by the growth of instabilities: 
\begin{equation}
\delta B (t) \approx \sqrt{4 \pi \frac{\xi_{\rm CR} \rho (t) v_{\rm s}^3(t)}{\Lambda c}}
\label{eq:B1}
\end{equation}
where $\xi_{\rm CR}$ is the CR efficiency (i.e., the fraction of the shock ram pressure converted into CRs at the shock through DSA), $\rho$ the density upstream the shock front, and $\Lambda = \ln(p_{\rm max}/mc)$, and the slope of accelerated particles is assumed to be $\propto p^{-4}$. 
Equation~\eqref{eq:B1} leads to typical values: 
\begin{equation}
\delta B (t) \approx 2 \left( \frac{\xi_{\rm CR}}{0.1} \right)^{1/2} \left( \frac{v_{\rm sh}(t)}{1000 \; \text{km/s}}\right)^{3/2} \left( \frac{n}{1~\text{cm}^{-3}}\right)^{1/2} \mu \text{G}
\label{eq:B1}
\end{equation}

Thus, for shock velocities of a few thousand km/s, and density sufficiently high, which is typical for instance of the shock launched by the explosion of core-collapse supernovae exploding in the dense wind of a late sequence massive star, values of few hundred $\upmu G$ can be reached. 

The growth of nonresonant instabilities drives the amplification of the magnetic field and thus allows to reach higher values for the maximum energy of accelerated particles. Refs. \citep{bell2013,schure2013} discussed that this maximum energy can be estimated by considering that the growth of instabilities saturates when the amplification reaches a number ${\cal N}$ of $e$-folding. For the maximum growth rate $\gamma_{\rm max}$, corresponding to wave numbers $k$, this condition reads: $\gamma_{\rm max} t \approx {\cal N}$. The value of ${\cal N}$ is not well constrained, but arguments in favor of $ 3 \leq {\cal N} \leq 10$ have been made~\citep{bell2013}. The corresponding maximum energy reads: 
\begin{equation}
E_{\rm max}(t) \approx \frac{1}{{\cal N}} e \sqrt{4 \pi \rho(t)} \frac{\xi_{\rm CR} v_{\rm sh}^3(t) t}{c \Lambda} 
\label{eq:Emax1}
\end{equation}

Expliciting Equation~\eqref{eq:Emax1} with numerical values: 
\begin{equation}
\label{eq:Emax1}
E_{\rm max}(t) \approx 1 \left(\frac{\xi_{\rm CR}}{0.1} \right) \left( \frac{v_{\rm sh}(t)}{5000~\text{km/s}} \right)^3 \left( \frac{t}{100~\text{year}} \right) \left( \frac{n}{10~\text{cm}^{-3}}\right)^{1/2} \text{PeV}
\end{equation}
where ${\cal N}=5$. This shows that in order to reach the PeV range at times of the order of the century, relatively high shock velocities and density are required. However, a careful calculation including the precise quantities associated to typical SNRs shows that in most of the evolution of  typical SNRs  - in most of the free expansion phase, and in the adiabatic Sedov-Taylor phase - the PeV range cannot be attained. Indeed, in SNRs from thermonuclear supernovae, high shock velocities $\gtrsim$5000 km/s can be observed but usually associated to an ISM of the order of $\sim$1 cm$^{-3}$. On the order hand, for SNRs from core-collpase SNe (CCSNe), which correspond to the bulk of SNe, the higher densities provided by the dense wind of the late sequence progenitor star can be as high as $\sim$$10^{3}$--$10^{4}$ cm$^{-3}$ (decreasing with $\propto r^{-2}$) but thereby substantially decreasing the shock velocity below $\lesssim$500 km/s. It therefore seems that only SNRs from peculiar SNe, sufficiently energetic and/or exploding in dense winds are capable of efficiently producing PeV particles~\citep{schure2014}. Moreover, the duration of the pevatron phase of these SNRs appears to be rather limited. By estimating the total content of protons produced by typical SNRs and peculiar pevatron SNe, it has been possible to use the local CR proton spectrum to constrain the rate of Galactic SNR pevatron to a fraction of the Galactic SN rate $\lesssim$3--4\% $\nu_{\rm SN} \approx 0.1/$century~\citep{cristofari2020}. Moreover, since these peculiar SNRs are typically pevatrons on timescales of $\lesssim$ one century, this corresponds to one SNR pevatron active for the order of $\sim$1 century every few \mbox{$\sim$10,000 years}, drastically limiting the chances of catching an active SNR pevatron. 
In other words, it could be that SNRs are indeed producing the bulk of CRs, but that the short duration of the pevatron phase prevent us from seeing one in activity. 

\subsection{Supernovae}
Discussions on magnetic field amplification, e.g., through the growth of nonresonant streaming instabilities, lead us to the idea that the maximum energy of accelerated particles scales in time as $\propto t v_{\rm sh}^3(t) \rho^{1/2}(t)$. This corresponds, for CCSNe in which the velocity is typically in the early stage (free expansion phase) $v_{\rm sh} \propto t^{-1/8}$ in a wind of profile $n \propto r^{-2}$ to $E_{\rm max} \propto t^{-1/4}$. In the case of a thermonuclear SNe expanding in a uniform ISM, $v_{\rm sh} \propto t^{-3/7}$ in the free expansion phase, thus $E_{\rm max}\propto t^{-2/7}$. This indicates that the maximum energy is reached at the earliest time and supports the idea that if some SNRs are pevatrons, they most likely must be pevatron when they are the youngest. Since the duration of the potential pevatron phase is virtually unknown, some authors have investigated the possibility that PeV acceleration would especially be efficient during a few days/month/years after the explosion of SNe~\citep{tatischeff2009,marcowith2014,murase2014,giacinti2015,zirakashvili2016,petropoulou2017,bykov2018,bykov2018b,marcowith2018,tamborra2018,murase2019,fang2019}. Consequently, such efficient particle acceleration could lead to a considerable emission of gamma rays from the GeV to the multi-TeV range, a priori mostly through hadronic interactions because of the high density of the circumstellar environment in which CCSNe explode. Of course, in our Galaxy, the problem of low number of events remain, with a SN rate typically inferred $\sim$3/century, but it has been argued that the detection in the gamma-ray domain of close-by extragalactic SNe could help to study the acceleration up the PeV range in the first days after the explosion of the SN (in other words, extremely young SNRs). 

The possibility for instruments to detect such SNe is essentially limited by a physical process that can degrade the gamma-ray signal: the two-photon interactions in which a gamma-ray photon interacts with a low energy photon from the SN photosphere, producing an electron-positron pair, can potentially degrade the gamma-ray by several orders of magnitude. In the case of luminous supernovae SN1993J, it has been shown that a detection of gamma rays from extragalactic energetic SNe within a few Mpc is possible with next-generation instruments~\citep{cristofari2020b}.

\subsection{Archeology of Pevatrons}
Whether it is SNe, or SNRs, or even other astrophysical objects, since the duration of pevatron phase is expected to be limited (month, years or few centuries), and that together with the low rate of these objects, we might be in a situation where the number of active Galactic proton pevatrons is low. See e.g., discussion above, on the number of SNR pevatrons from energetic SNe expected to be $\lesssim$0.1/century typically active for a duration of the order of $\sim$1 century. This would make it possible for us to be in a situation where SNRs do produce the bulk of Galactic CRs, but where we currently do not have any active proton SNR pevatron in the Galaxy. 
In such a situation, we need alternative ways to probe the past activity of pevatrons, and therefore attempt to do archaeology of pevatrons~\endnote{A term borrowed to J. Holder.}. The signature of the extinct pevatron could for instance be found through deeper studies of SNR shocks (e.g., extrapolating backwards the shock dynamics), by probing the surrounding molecular clouds (MC) of the pevatron candidates, for instance through gamma rays observations of MCs in the vicinty of pevatron candidates~\citep{gabici2009,baghmanyan2020,mitchell2021}, or through the study of the radiation of secondary particles produced around pevatrons~\citep{celli2020}.

\section{The Detection of Galactic Pevatrons with Gamma Rays}

\subsection{The Crab Nebula}
The Crab Nebula is probably one of the best studied objects of the Universe: its emission has been monitored over several decades and accross the entire accessible electromagnetic spectrum~\citep{meyer2010}. 
Extensive observations of the Crab Nebula have revealed variability in the gamma-ray domain~\citep{balbo2011,striani2013,aleksic2011}. These \textit{flares}---typically lasting for several hours to few days---have often been interpreted as the result of efficient acceleration of electrons up to the PeV range, emitting gamma rays through synchrotron mechanism. Discussions on the acceleration mechanisms usually invoke magnetic reconnection around the Crab pulsar and shock acceleration at the pulsar-wind termination shocks~\citep{uzdensky2011,arons2012,bykov2012,lyutikov2012,cerutti2014,lyutikov2016,giacinti2018,khangulyan2020}. Recently, the LHAASO instrument reported on the detection of gamma rays above \mbox{$\gtrsim$1 PeV}, additional confirmation that an electron pevatron is located within the Crab Nebulae~\citep{crab_lhaaso,cao2021}. This means that, unexpectedly, the first detection of a pevatron corresponded to an electron--pevatron, a reminder that the finding of a pevatron does not immediately solve all problems of CR physics.

Currently, the mechanism involved in electron acceleration at the Crab Nebula are still not well understood, which makes the case of the Crab Nebula even more remarkable. Several mechanisms have been proposed, such as for instance: (a) DSA at the termination shock formed between the pulsar relativistic wind and the circum-pulsar medium; (b)~driven magnetic reconnection of the alternating field lines compressed at the shock; (c)~resonant cyclotron absorption in ion-doped plasma, in which electrons and positrons absorbe energy emitted by ions in the plasma~(see e.g., for a short and pedagogical discussion~\citep{amato2020}). Theoretical works, through Magnetohydrodynamics (MHD) and particle-in-cell (PIC) simulations are helping to advance our understanding on particle acceleration around pulsars.

\subsection{The First Galactic Pevatron Detected through VHE Gamma Rays}
While the multiple observations in VHE range with IACTs of SNR shells were revealing that all known SNRs seem to not be pevatrons, observations with H.E.S.S. were objectifying that gamma rays up to $\sim$100 TeV seem to originate from the Galactic center region around Sagittarius A*, a source called HESS J1745-290, thereby indicating the presence of a pevatron in the Galactic center. No SNR seem to be associated to this emission, making the first catch of a pevatron even more startling than expecting. As previously mentioned, the 100 TeV range is not directly accessible to H.E.S.S., but the observations were capable of revealing a hard spectrum with little suppression up to 10--20 TeV, strongly suggesting that 100 TeV gamma rays are produced~\citep{HESSpevatron}. This result as been confirmed by other instruments such as VERITAS~\citep{adams2021}. Various interpretations of the gamma-ray emission of this pevatron candidate have been proposed, including e.g., supernovae and clusters of massive stars~\citep{bykov2018,biermann2018}, millisecond pulsars~\citep{hooper2018,hooper2018b,guepin2018}, or particle acceleration from the supermassive black hole~\citep{fujita2017,istomin2020}.

\subsection{A Population of Galactic Pevatrons}
In the recent years, refined observations with various techniques, such as IACTs (e.g., H.E.S.S.~\citep{HESS2006}, VERITAS~\citep{VERITAS}, MAGIC~\citep{MAGIC})~(see e.g., for reviews \citep{hillas2013,rieger2013,caraveo2020}), shower front detector instruments (e.g., HAWC~\citep{HAWC2017}, Tibet AS$\gamma$~\citep{Tibet2019}, LHAASO~\citep{bai2019}), and systematic survey of the sky, have helped discover several Galactic pevatron candidates.
The HAWC observatory has for instance reported on the detection of at least ten sources with emission above $\gtrsim$50 TeV of likely Galactic origin~\citep{abeysekara2019,albert2020,albert2020}. The case of J2227+610 is especially exciting as a new potential Galactic pevatron~\citep{albert2020}, since it has also been detected with Tibet AS$\gamma$~\citep{2021tibet}, and its association with the SNR G106+2.7. Arguments in favor of the acceleration of PeV particles from the pulsar wind nebula VER J2227+608 through gamma-ray observations with Fermi-LAT had already been made~\citep{xin2019}, and make this case of special interest, although the situation remains quite unclear due to the complexity of the region~\citep{pineault2000,kothes2001}. 

Finally, with the detection of gamma rays of energy $\gtrsim$100 TeV by LHAASO from a dozen of sources that are likely located in the Galaxy, another important milestone has been reached: the first detection of a population of pevatrons~\citep{cao2021}. So far, the spectral shape of most of these sources is still not well established, which makes it difficult to clearly understand whether the origin is leptonic or hadronic. Moreover, the source association remains a difficult task given the typical angular resolution of the order of $\sim$0.3$^{\circ}$ at 100~TeV, but these detections are still an important step in the hunt for Galactic pevatrons. 
In Table~\ref{tab:list} we give a quick overview of the known Galactic pevatron population as of May 2021, although this list is likely to be lengthened soon due to ongoing surveys. 

\end{paracol}
\nointerlineskip
\begin{specialtable}[H]
\widetable
\caption{List of known Galactic pevatrons as of May 2021. This list is likely to be lengthened soon due to active ongoing detection campaigns.}
\tabcolsep=0.436cm
\begin{tabular}{lcr}
\toprule
\textbf{Source} & \textbf{Possible Association} & \textbf{Reference} \\
  \midrule			
  HESS J1745-290 & Sagittarius A*/Galactic center& \citep{HESSpevatron} \\ 
  Crab/LHAASO J0534+2202 & PSR J0534+2200& \citep{balbo2011,aleksic2011,Tibet2019,cao2021} \\
  LHAASO J1825-1326/2HWC J1825-134  & PSR J1826-1334/PSR J1826-1256 & \citep{albert2021,cao2021} \\
   LHAASO J1839-0545/2HWC 1837-065 & PSR J1837-0034/PSRJ1838-0537&  \citep{abeysekara2019,cao2021} \\
      LHAASO J1843-0338/2HWC J1844-032& SNR G.28.6-0.1&  \citep{abeysekara2019,cao2021} \\
            LHAASO J1849-0003 & PSR J1849-0001/W43&  \citep{cao2021} \\
            LHAASO J1908+0621/MGRO 1908+06/  & SNR G40.5-0.5/PSR 1907+0602/PSR 1907+0631&  \citep{abeysekara2019,cao2021} \\
                     2HWC 1908+063 & & \\
            LHAASO J1929+1745 & PSR J1928+1746/PSR1930+1852/SNR G54.1+0.3&  \citep{cao2021} \\
                        LHAASO J1956+2845 & PSR J1958+2846/SNR G66.0-0.0&  \citep{cao2021} \\
                        LHAASO J2018+3651 & PSR J2021+3651/Sh 2-104 (HII/YMC)&  \citep{cao2021} \\
                       HWC J2019+368  & & \citep{abeysekara2019} \\ 
                        LHAASO J2032+4102/2HWC J2031+415 & Cygnus OB2/PSR 2032+4127/SNR G79.8+1.2&  \citep{abeysekara2021,cao2021} \\
                        LHAASO J2108+5157 &&  \citep{cao2021} \\
                        LHAASO J2226+6057 & SNR G106.3+2.7/PSR J2229+6114&  \citep{2021tibet,cao2021} \\
                          HESS J1702-420A &  SNR G344.7-0.1/PSR J1702-4128& \citep{giunti2021,giunti2021_B} \\
  \bottomrule  
\end{tabular}
\label{tab:list}
\end{specialtable}
\begin{paracol}{2}
\switchcolumn

\vspace{-10pt}
\section{All Kinds of Pevatrons: Pulsars, Massive Stars, Stellar Clusters \& Superbubbles}
\subsection{Pulsars and Surroundings}
The Crab Nebula, as mentioned above, is a clearly accepted example of pevatron, where the gamma-ray flares reveal the acceleration of ultrarelativistic electrons, injected from the pulsar and are energized to at least the PeV range. Altough it is possible that protons are also accelerated, the Crab is a clear example of an electron pevatron. It illustrates the possibility for pulsars, and systems hosting pulsars to be efficient particle accelerator. The detection of several halos of TeV gamma rays, often referred to as \textit{TeV halos} is connected to this issue of particle acceleration around pulsars~\citep{sudoh2019,giacinti2020,crab_lhaaso}. 

\subsection{Massive Stars, OB Associations and Stellar Clusters}
Aside from the fact that most SNRs seem to not be pevatrons, other issues remain: for example, the spectra of accelerated particles at SNR shocks, the spectra of particles injected in the ISM, or the electron-to-proton ratio accelerated and injected in the ISM. These tensions have motivated the search of potential alternative accelerators: massive stars, and their dense winds, leading to the formation of a strong shock at the termination of the shocks, were for instance proposed as Galactic CR factories~\citep{voelk1982,cesarsky1983,cesarsky1983b}. Moreover, as massive stars tend to form in clusters, it has been proposed that the shock created by the collective effects of the winds of individual stars could accelerate protons above the PeV range, and contribute to the production of Galactic CRs~\citep{stevens2003,morlino2021}. For a review on particle acceleration from star forming regions we refer the reader to~\citet{bykov2020}. 

In addition to theoretical arguments, the detection of gamma rays from star clusters has been a direct confirmation that efficient particle acceleration is taking place: Westerlund 1 and Westerlund 2 detected by H.E.S.S.~\citep{abramowski2012,yang2018}. 
Moreover, spatially resolved observations of stellar clusters have revealed that the distribution of particles from massive stars seem to scale as the inverse of the distance away from the sources, which is an indication of steady production and injection of particles in the ISM. This is thus supporting the idea that these objects could be prominent sources of CRs~\citep{aharonian2019,saha2020,sun2020}. Futhermore, the detection of a population of at least 12 sources emitting gamma rays above $\gtrsim$100 TeV by the new instrument LHAASO, testifies of the detection of at least 12 pevatrons of Galactic origin, at reported in Table~\ref{tab:list}. Although it is yet difficult to draw any firm conclusion on the association to known objects, most of these sources seem to not be associated to SNR shocks. On the other hand, some sources, as for example, J2032+4102 could potentially be associated to massive stars. Precise measurements of the differential gamma-ray spectra of these sources are needed to understand if the emission is due to the acceleration of protons or electrons above the PeV range, and deeper studies of the spatial origin of this emission will help understand which astrophysical object are responsible for the production of these particles.

\subsection{Superbubbles}
In the line of star clusters, superbubbles, the giant cavities of hot and turbulent plasma inflated by repeated neighboring SNe explosion, have also been proposed as potential pevatrons. Superbubbles have especially been mentioned in the context of the search for the origin of CRs because they could potentially account for unexplained trends in the CR composition. Indeed, measurement of the CR spectrum have indicated that the abundance of volatile elements (e.g., N, Ne, Ar) seem to correlate with the atomic mass; a correlation that is not clearly found for refractory elements (e.g., Mg, Si, Fe)~(see \citep{tatischeff2018} for a review on the topic) . It has been shown that a mixture of material with primordial solar composition and material enriched by stellar outflows or ejecta could explain the observed trends~\citep{murphy2016}: such a mixed ISM can typically be found in superbubbles. 
Moreover, superbubbles might help overcome another identified problem: the problem of the abundance of Be and B. In metal poor halo stars, Be and B have been found to increase linearly with metallicity, rather than quadratically, which is expected in a standard scenario in which CRs are accelerated in a nonenriched ISM~\citep{parizot2000}. Superbubbles might help explain this linear dependency. This makes them interesting candidates in the CR origin problem, and as any good CR factory candidate, the question of particle acceleration up to the PeV range is central~\mbox{\citep{koo1992,koo1992b,bykov2001,parizot2004,gupta2018}}. Detection of gamma rays from the Cygnus X superbubble~\citep{ackermann2011} supports the need for further investigations with next-generation instruments optimized in the 100~TeV range. 
Finally, the detection of high-energy neutrinos from the Cygnus Cocoon region, in association with gamma rays, comes as direct indication that efficient acceleration of hadrons is taking place, and has been claimed to be a clear evidence that a proton pevatron is located in the Cygnus region~\citep{dzhappuev2021}. 

\subsection{Other Candidates}
In the list of potential proton pevatrons, and thus possible contributors to CR proton spectrum, we must also mention other astrophysical objects, such as, for example, low-luminosity black holes binaries and their magnetically arrested disks~\citep{kimura2021}, or direct acceleration of protons in the magnetospheres of pulsar~\citep{guepin2020}. Although we only briefly mention these recently proposed scenarios here, they clearly deserve full attention and must be kept in mind in the hunt for pevatrons.

\subsection{Superpevatrons}
The improved performance of instruments have made possible the detection of gamma rays of energies even above 100 TeV. For instance, 
HAWC has reported on the detection of gamma rays around $\sim$200 TeV from J1825-134~\citep{albert2020} and TibetAS$\gamma$ on the detection of gamma rays $\gtrsim$100 TeV potentially associated to the region around SNR G106+2.7 or PSR J2229+6114, as previously discussed~\citep{2021tibet}, and the LHAASO collaboration has reported on the detection of gamma rays of $\gtrsim$1 PeV around the source J2032+4102, or from the Crab Nebula~\citep{crab_lhaaso}. Should these gamma rays be due to hadronic interactions, this would indicate the acceleration of $\gtrsim$10 PeV CR hadrons, and thus point to Galactic superpevatrons. These superpevatrons might thus play a role in shaping the CR spectrum above the \textit{knee}, and open many questions on the origin of CRs above the knee, and on the transition between CRs of Galactic and extragalactic origin.

\section{Gamma Rays: Limitations and Hopes for the Future}

Gamma-ray instruments have now clearly demonstrated their potential in the search for pevatrons. It remains that several or most of the aspects of particle acceleration up to the PeV range have yet failed to be understood through observations in the gamma-ray range: the astrophysical site(s), the physical mechanism(s), the duration of the pevatron phase(s). In order to move forward on these issues, it is essential to progress on the spatial, time and spectral resolution. This statement is quite general in Astronomy, since improving performance is always a strong motivation behind new generations of instruments at any wavelength: in the case of gamma-ray instruments and the search for pevatrons, it is now essential in VHE and UHE domain.
It is worth mentioning typical orders of magnitude that are determining in the context of the hunt for pevatrons. First, the time evolution of the sensitivity of gamma-ray instruments makes it usual to target sources for at least several tens of minutes, and usual to need $\sim$tens of hours, which naturally limits the amount of observing time available per source. In turns, this often prevents multiplying observations of a target of interest on monthly scales, and thus makes it challenging to probe the time dependence of the gamma-ray emission. Second, the angular resolution, even in the most favorable case of next generation IACTs (such as CTA)~\citep{CTAscience}, is at best of the order of a few arcmin at $\sim$100 TeV and $\sim$10 arcmin at 100 GeV, which limits the possibility of probing extended emission even around Galactic accelerators. This is problematic, for instance to address the question of the escape of particles from their accelerators, or to study complex regions around pulsars (e.g., TeV halos)~\citep{hooper2021}. 

Finally, the spectral perfomance. Again, even in the favorable case of IACTs, obtaining a differential spectrum from a point source requires a significant amount of observing time (typically several hours), and the obtained spectrum is usually composed of $\sim$10--20 bins per decade at $\sim$1 TeV, with error bars increasing at high energy with the decreasing fluxes. 
A least two comments are in order. In the context of the search for pevatrons through gamma-ray observations, it is now clear that the shape of the spectrum above $50$ TeV is crucial (and especially above 100 TeV), if only to discriminate between the potential leptonic and hadronic origin. We have seen that with the remarkable performance of air shower instruments (such as LHAASO), integrated spectra can reveal the presence of pevatrons, but the characterization of their spectra is also essential: spectral index, existence of an exponential suppression, gradual change of slope.
These details imprinted in the gamma-ray spectrum can unveil precious information on the parent particle content. For instance, a sharp exponential cutoff in the 50 TeV range is expected in the case of parent electrons, due to the Klein--Nishina suppression, whereas a more gentle steepening could sign the presence of parent protons, or of synchrotron from electrons, as for example with the Crab gamma-ray spectrum and its gradual steepening over three energy decades \citep{crab_lhaaso}.
Moreover, differences in the initial spectra of parent protons and electrons could also imprint the gamma-ray spectra. For example, at SNR shocks, it has been shown that the high-energy suppresion for protons is often well approximated by an exponential $\propto \exp(-p/p_{\rm max})$ whereas for electrons, if the diffusion is Bohm-like and the maximum energy loss limited, the spectrum is expected to be suppressed as $\propto \exp(-\left(p/p_{\rm max}^e \right)^2 )$~\citep{zirakashvili2007,blasi2010}. 

Futhermore, it has been shown that the shape of the suppression at VHE energy can also vary for different accelerators, even considering only the acceleration of protons. This is for instance the case of stellar clusters. As detailled in~\citet{morlino2021}, which proposed a semianalytical description of the acceleration of particles at the wind termination shock of stellar clusters, the geometry, the diffusion regime both inside the stellar cluster wind and outside the termination shock can shape the suppression at the highest energies. In principle, the study of the gamma-ray spectral index at the highest energy could therefore offer precious indications on physical parameters at accelerators of interest, although, of course, such study will remain extremely challenging even for next generation IACTs. 

Eventually, let us observe that gamma-ray astronomers have taken the habit of trying to fit differential spectra with power laws, broken power laws, or power law exponentially suppressed, or other usual simple functions. This approach is physically reasonable, effective, and useful, but it can sometimes become decorrelated from any underlying assumption on the physical mechanism involved. Moreover, in many cases, the quality of the data after high-level analysis is still limited and allows for some freedom in the fitting. In some cases, one can then wonder if a given function providing a decent fit to a set of data, is really more motivated than another, and it has to be kept in mind that the fitting with usual function might be artificial and lead to loss of information. As mentioned above in the case of stellar cluster, at VHE, deviations from a clean exponential cutoff could sign different physical aspects, and it is especially important to keep this point in mind for future investigations of the gamma-ray spectra in the $\gtrsim$50--100 TeV range. Indeed a wide variety of spectra could be revealed in this range, from steep to exponentially suppressed, or modified exponentials, including gradual changes of index or even bumps and features, holding valuable information. 

In addition to the gamma-ray domain, it is essential to mention observations at other wavelengths useful in the pevatron search, already down to the radio domain, where for synchrotron emission of accelerated electrons or of secondary particles can help probe the processes at stake~\citep{balagopal2018,cristofari2018,celli2020,ge2021}. Multimessenger observations (especially neutrinos) are also encouraging~\citep{gabici2007,gonzalez2009,gladilin2018}~(see e.g. \citep{anchordoqui2014,auger2019,ligoreview} for reviews).

\section{Open Questions and the Hunt in the Coming Years}
The recent detection of several Galactic pevatrons with various instruments operating the gamma-ray domain is gripping and might help us close in on the origin of Galactic CRs. Although it has often been claimed that the detection of 100 TeV gamma rays would be the unequivocal proof of the acceleration of PeV protons, and thus almost straightforwardly uncover the sources of Galactic CRs, it is now clear that things are a bit more abstruse.
The detection of Galactic pevatrons has yet maybe brought more questions than answers. Among these questions:
\begin{enumerate}
\item Is the gamma-ray emission from Galactic pevatrons due to the acceleration of electrons or protons?
\item What mechanism(s) drive(s) the acceleration of particle up to the PeV range? 
\item What astrophysical object(s) is (are) responsible for the acceleration of PeV protons (and thus likely candidate for the origin of CRs)? 
\item How long do the pevatron phases last? 
\item When do PeV CRs escape their sources? What is the distribution of particles around the pevatrons?
\item What is the population and distribution of Galactic proton and electron pevatrons? 
\item If the proton knee is, as suggested by some experimental results, below 1 PeV (\mbox{$\sim$0.7 PeV}), how does this affect the search for the sources of Galactic CRs?
\item Which objects are superpevatrons? How do they contribute to the CR spectrum? 

\end{enumerate}

With the detection of several Galactic pevatrons, whose astrophysical nature and spectral properties have yet to be understood, it is now clear that finding $\gtrsim$100 TeV gamma rays, and catching a pevatron, is not enough to solve the problem of the origin of Galactic CRs.
Sources of Galactic CRs must necessary be pevatrons, but this condition is not sufficient. In order to win the title of ``main source(s) of Galactic CRs'', an astrophysical site will not only have to be a pevatron, but also to answer the different issues currently faced by SNRs such as: spectrum of particle injected in the ISM, chemical composition of CRs ($^{22}$Ne/$^{22}$Ne ratio), fondamental physical process driving the acceleration of particles and magnetic field amplification, etc. 
Moreover, the role of the different detected pevatrons in the production of CRs and their contribution to the LIS will have to be understood.
Fundamentally, one also has to prepared to the idea that not only one class (or ``type'') of source(s) produce the bulk of Galactic CRs, but that the local CR spectrum might in fact the result of a complex intertwining of many sources, e.g., acceleration and injection at very high energy from some sources for a limited amount of time, reacceleration at other sources, etc., that would finally produce the measured local CR spectrum. 
The study of pevatrons will undoubtedly help to better understand the fundamental physical processes at stake (growth of instabilities, escape from sources, content injected in the ISM from sources), but will likely not be sufficient to answer all questions opened by CR physics. Complementary studies of the propagation of CRs in the Galaxy, or CR composition are vital to provide a complete understanding of the local CR spectrum. 

In the coming years, multiwavelength and multimessengers observations will also be essential to in order to progress on the question of origin of CRs. The gamma-ray range itself is expected to play a central role to probe the content of accelerated particles, and study the astrophysical objects involved. The instruments already operating in the VHE range, such as H.E.S.S., Veritas, MAGIC, LHAASO, HAWC, and the next-generation instruments, such as CTA~\citep{CTAscience}, SWGO~\citep{SWGO}, with constantly improving spectral energy resolution and angular resolution, are needed to better understand the gamma-ray spectra in the $\sim$100 TeV range, the emitting gamma-ray regions, and the distribution of particles around Galactic pevatrons. 

\vspace{6pt}




\funding{{This research received no external funding.}}

\institutionalreview{{Not applicable}}

\informedconsent{{Not applicable}}

\dataavailability{{Not applicable}} 

\acknowledgments{PC warmly thank F. Aharonian, P. Blasi, S. Gabici, J. Holder, A. Marcowith, P. Martin, G. Morlino, E. Peretti, M. Renaud, H. Sol, and the entire LUTH PHE team for stimulating discussions.}

\conflictsofinterest{{The authors declare no conflict of interest.}} 

\begin{adjustwidth}{-4.6cm}{0cm}

\printendnotes[custom]

\end{adjustwidth}
\end{paracol}

\reftitle{{References} 
}

\end{document}